# A Modified DNA Computing Approach to Tackle the Exponential Solution Space of the Graph Coloring Problem

Ramin Maazallahi[1] and Aliakbar Niknafs[2]

[1]Department of Computer Engineering, Shahid Bahonar University of Kerman, Iran
`ramin.mz66@gmail.com`
[2]Department of Computer Engineering, Shahid Bahonar University of Kerman, Iran
`niknafs@uk.ac.ir`

*ABSTRACT*

*Although it has been evidenced that DNA computing is able to solve the graph coloring problem in a polynomial time complexity, but the exponential solution space is still a restrictive factor in applying this technique for solving really large problems. In this paper a modified DNA computing approach based on Adleman-Lipton model is proposed which tackles the mentioned restriction by coloring the vertices one by one. In each step, it expands the DNA strands encoding promising solutions and discards those which encode infeasible ones. A sample graph is colored by simulating the proposed approach and shows a notable reduction in the number of DNA strands used.*

*KEYWORDS*

*DNA Computing, Graph Coloring, Exponential Solution Space, Adleman-Lipton, NP-Complete*

## 1. INTRODUCTION

The graph coloring problem is defined as assigning colors to the vertices of a graph such a way that no two adjacent vertices have the same color and a minimum number of colors are used. This problem has many applications such as fault diagnosis [1], functional compression [2], broadcast scheduling [3], resource allocation [4] and biological networks [5].

The graph coloring problem is one of NP-complete problems, where the number of possible color assignments increases exponentially with respect to the number of vertices. This makes the solution space of the problem exponential; therefore a brute force algorithm fails to check all the assignments to find the solutions. In situation when the exact algorithms are not applicable, heuristic methods are used which give satisfactory solutions but do not always guaranty the optimality of solutions.

The largest saturation degree (DSATUR) [6] and the recursive largest first (RLF) [7] are of the first heuristic approaches proposed to solve the graph coloring problem. They operate greedy and color the vertices of the graph one by one. DSATUR algorithm evaluates a degree of saturation for each vertex of the graph and starts by assigning the first color to a vertex of maximal degree. The degree of saturation is reevaluated for the remaining vertices and the next vertex to be colored is selected. The RLF algorithm selects the vertex with the largest number of uncolored neighbors and colors that vertex. It then removes that vertex and its neighbors and selects the next uncolored vertex. The process is repeated until all of the vertices of the graph are colored. Although these approaches are fast, but they usually yield the locally optimal solution.

Genetic algorithm (GA) as a search heuristic which mimics the process of natural evolution has been used in [8] to solve the graph coloring problem. The authors claim that GA outperforms DSATUR on the hardest graphs, although for small and large graphs DSATUR performs better. Also the time complexity of GA is lower than DSATUR.

In some researches the use of local search algorithms have been proposed for solving the graph coloring problem. Some examples are Tabu Search [9], Simulated Annealing [10] and Variable Space Search [11]. Also neural network approach for graph coloring is proposed by Rahman [12]. Dorrigiv [13] proposes a hybrid algorithm based on Artificial Bee Colony (ABC) algorithm and RLF algorithm. In their approach, first the ABC algorithm is used to generate a sequence of nodes and then the RLF algorithm is applied for coloring the vertices of the graph.

In recent decades, the efficiency of using DNA computing for solving NP-complete problems has been demonstrated (see for example [14-17]). DNA computing has the advantage of searching the whole exponential solution space of the problem by applying polynomial number of biological operations with respect to the problem's input size. DNA computing has already been used for solving the graph coloring problem in some researches [18-21]; But the main problem of applying DNA computing for large graphs is the exponential number of DNA strands needed for computation.

In this paper we propose a modified DNA computing approach for solving the graph coloring problem which does not need the initial exponential space of solutions. We start with the first vertex and generate DNA strands coloring that vertex correctly. Then we add the second vertex and expand the previous DNA strands to correctly color both the vertices. This process is repeated until the whole of vertices are colored. Expanding the promising solutions and discarding the infeasible ones, makes this possible to solve the graph coloring problem for larger graphs than before.

The rest of this paper is as follows. Section 2 briefly describes DNA computing and the Adleman-Lipton model which has been used in this paper. In Section 3, we present our approach in details and simulate it for solving a sample graph in Section 4. Finally, in Section 5 we draw conclusions and present suggestions for future work.

## 2. DNA COMPUTING

DNA (deoxyribonucleic acid) is a polymer made up of a linear arrangement of monomers called nucleotides. Distinct nucleotides are only detected with their bases, where there are four different bases known as *Adenine*, *Thymine*, *Cytosine* and *Guanine*, abbreviated as A, T, C and G respectively. Nucleotides are simply referred to as A, T, C and G nucleotides, depending on the kind of base that they have. Therefore we can represent DNA strands as strings over the alphabet {A,T,C,G}.

DNA computing is a novel computational paradigm that uses DNA molecules for storing information and biological operations, such as *polymerase chain reaction* (PCR), *ligation* and *gel electrophoresis*, for acting on the information stored. Considering that a test tube can contain up to $10^{18}$ DNA molecules [22] and a biological operation will act on all of the molecules simultaneously, the power of DNA computing becomes apparent. It resembles a system with $10^{18}$ processors running in parallel. In every DNA computing experiment, a test tube is used for storing the initial DNA strands which each of the strands encodes a candidate solution to the problem. The computation then carries on by applying a sequence of biological operations which filter out infeasible candidate solutions.

The vast parallel processing ability of DNA molecules makes DNA computing a powerful tool for solving intractable problems. Adleman [23] was the first who used DNA molecules and biological operations for solving an instance of the Hamiltonian path problem in laboratory. Following Adleman's success, Lipton used the same technique to solve the satisfiability problem [24]. The biological operations proposed by Adleman and Lipton constructed the Adleman-Lipton model. Several other models have also been proposed for DNA computing, such as the sticker model [25], parallel filtering model [26], split-and-merge filtering [27] and filtering-by-blocking model [28].

The Adleman-Lipton model is constructed upon the following biological operations:

- **Append (T, S):** Given a tube T and a string S, this operation appends S to the end of all the DNA strands of T.
- **Copy (T, $T_1$, $T_2$, …, $T_n$):** Given a tube T, this operation creates tubes $T_1$ to $T_n$ which are identical copies of T. ($T_1$ to $T_n$ will have the same DNA strands as T.)
- **Merge (T, $T_1$, $T_2$, …, $T_n$):** Given tubes $T_1$ to $T_n$, this operation pours the contents of $T_1$ to $T_n$ into tube T; therefore T contains all the DNA strands of $T_1$ to $T_n$.
- **Extract (T, S, T+, T-):** Given a tube T and a string S, this operation produces two tubes T+ and T- as follows: All the DNA strands of T having S in their sequence are extracted and poured into T+ and the remained DNA strands are poured into T-.
- **Detect (T):** Given a tube T, this operation returns true if there is at least one DNA strand in T, otherwise it returns false.
- **Discard (T):** Given a tube T, this operation ignores T.

## 3. PROPOSED APPROACH

In this section, we propose our approach for coloring the graph by three colors (3-vertex coloring problem) and the generalization of our approach for *k*-vertex coloring problem is straightforward.

Let $G = (V, E)$ be an undirected graph. $V = \{v_1, v_2, ..., v_n\}$ is the set of vertices and $E = \{e_1, e_2, ..., e_m\}$ is the set of edges. For each $e_i \in E$, $e_i = (v_j, v_k)$ where $v_j$ and $v_k$ are two vertices from $V$ and are said to be adjacent vertices. The goal is to assign three colors of *red*, *green* and *blue* to the vertices in $V$, such a way that no two adjacent vertices share the same color.

To solve the problem by DNA computing, we need three distinct DNA strands for encoding the three possible color assignments for each vertex. Let $r_i$ be a DNA strand encoding the red color for $i^{th}$ vertex, $v_i$. In the same way, $g_i$ and $b_i$ are DNA strands which encode the *green* and *blue* colors for $v_i$, respectively. If these strands are concatenated to each other, they form longer DNA strands encoding a candidate solution for the graph coloring problem. For a graph with $n$ number of vertices, a candidate solution is represented by a DNA strand as follows:

$$c_1 c_2 ... c_n \qquad (1)$$

where $c_i \in \{r_i, g_i, b_i\}$ is a DNA strand encoding the color assigned to $v_i$. In all DNA computing experiments, we start with a test tube containing all possible candidate solutions to the problem and then filter out infeasible solutions to get a test tube of strands, each encoding a possible solution. But this approach fails for large problems, because the number of DNA strands in the initial test tube grows exponentially with respect to the input size of the problem. We propose to make the solution space of the problem, step by step, to avoid the huge number of DNA strands needed.

We color the vertices of the graph, one by one. Initially DNA strands are generated to color only the first vertex of the graph. Then these strands are expanded to color the second vertex too. In this step all the DNA strands that encode an infeasible solution to the problem are poured out and are not expanded in the rest of the algorithm. The remaining DNA strands are then expanded to color the third vertex too. Again infeasible solutions are discarded to avoid exponential growth of solution space. This process is repeated until all of the vertices of the graph are colored. Figure 1 shows our proposed algorithm based on Adleman-Lipton model.

```
1:  Initialize T_0 as an empty tube
2:  for i=1 to n
3:      Copy (T_0, T_red, T_green, T_blue)
4:      Append (T_red, r_i)
5:      Append (T_green, g_i)
6:      Append (T_blue, b_i)
7:      for j=1 to i-1
8:          if v_i and v_j are adjacent vertices
9:              Extract (T_red, r_j, T_red_bad, T_red)
10:             Extract (T_green, g_j, T_green_bad, T_green)
11:             Extract (T_blue, b_j, T_blue_bad, T_blue)
12:         end if
13:     end for
14:     Merge (T_0, T_red, T_green, T_blue)
15:     Discard (T_red_bad)
16:     Discard (T_green_bad)
17:     Discard (T_blue_bad)
18: end for
```

Figure 1. Proposed algorithm

We start with an empty tube denoted by $T_0$. In lines 3 to 6, $v_i$ is colored by *red*, *green* and *blue* colors. Then the inner for loop extracts DNA strands encoding an infeasible solution until now by checking the previously colored vertices. If $v_j$ is adjacent to $v_i$, then $v_j$ and $v_i$ cannot have the same color. So the DNA strands that assign *red* color to both of these vertices are extracted and poured into tube $T_{red\_bad}$ to denote illegal color assignments (line 9). Lines 10 and 11 do the same thing for *green* and *blue* colors. After the inner loop finishes, the DNA strands remained in $T_{red}$, $T_{green}$ and $T_{blue}$ are legal color assignments. These tubes are merged into tube $T_0$ in line 14. By finishing the $i^{th}$ cycle of the outer for loop, $T_0$ contains DNA strands that correctly color all the vertices from $v_1$ to $v_i$. The illegal color assignments are discarded in lines 15 to 17 to avoid exponential growth of solution space. After the algorithm finishes, $T_0$ contains DNA strands which correctly color all the vertices from $v_1$ to $v_n$.

## 4. EXPERIMENTAL RESULTS

We have verified the correctness of our approach through extensive simulation results but to have a comparison with previous works, we present the results for the graph used in [21]. Figure 2 shows this graph which consists of 12 vertices. The DNA strands used in our simulation are shown in Table 1.

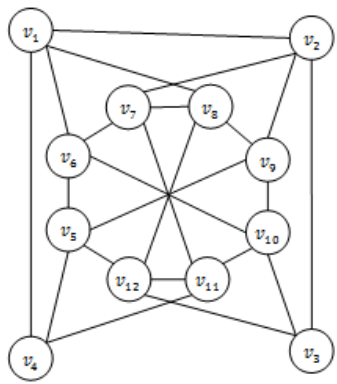

Figure 2. A sample graph with 12 vertices from [21]

Table 1. DNA strands used for solving the graph coloring problem shown in Figure 2

| Color | DNA strand | Color | DNA strand |
|---|---|---|---|
| $r_1$ | AAGGCAGGAACAGATCAACC | $r_7$ | TCGCTGCGATTCGATTTGTG |
| $g_1$ | CGTTCTAAATAGGGTCGTGT | $g_7$ | CCTCAGCGCCTCCGCGTAGC |
| $b_1$ | GATTAGACTTAGCTCGTCCG | $b_7$ | GCTCATCGTCGAAGCGTAGA |
| $r_2$ | CCACAATGTTATAATACCAC | $r_8$ | GTTCAATCCTTGCAGCCTCG |
| $g_2$ | ATCTTAGCACGATTCTCCTG | $g_8$ | CGTATAGAGCTGCACCATAC |
| $b_2$ | GTATATTCAAGTCTCGAGCC | $b_8$ | CGCAGGCAATAAGGGATTTG |
| $r_3$ | TTTAGATGAAACTCGCGTTC | $r_9$ | CTCCGATTAATGCACATTTA |
| $g_3$ | TGGCACTCTTAAATCGAATA | $g_9$ | GTTCGCGGATAAGAAGTCGA |
| $b_3$ | TTGACAAGGAGGAGGATCCA | $b_9$ | GCGTCCTAGGATCGTTCATT |
| $r_4$ | TCGGGGTAAAGTGATTACTG | $r_{10}$ | TTCCCTTTCCGGACTCTTCG |
| $g_4$ | ACCGATCAGTAACTAAATTC | $g_{10}$ | GGCTACTTCTTGTTACTCCA |
| $b_4$ | CGATGAGCGCCCTGAGGGGC | $b_{10}$ | TAACTGAATCGTCCAATCAC |
| $r_5$ | CGCCGCGTAAGGAGCCCGGT | $r_{11}$ | CAAACTGCTACGTCGCCAAT |
| $g_5$ | ACTTATCTTATAAGCGCCGG | $g_{11}$ | GGCTCCGAAACGATGGAAGT |
| $b_5$ | GGTCCAGCCTAACTTTTCAT | $b_{11}$ | TTCTTGGGGCTTGGGCTATA |
| $r_6$ | ATCTTGACCGCCAATATAAG | $r_{12}$ | CTCACAGAATGCTGCGCAAA |
| $g_6$ | CCAATTGTGCCAGCACGTTA | $g_{12}$ | TAAATTTACTTCGGGACACC |
| $b_6$ | AGATACCCGTCTGGTTCACC | $b_{12}$ | TCTCAACAGCGTCTGGAAGT |

A typical DNA computing approach for coloring the graph shown in Figure 2 needs an initial test tube of $3^{12}$ different DNA strands to encode all the solution space; but in our approach the solution space was searched step by step and the number of different DNA strands in a test never exceeded 180.
The idea of reducing the number of DNA strands was already proposed by Xu et. al [21]. It is reported in [21] that based on their approach, the initial solution space included 238 DNA strands, so the effectiveness of our approach is demonstrated. This reduction in the number of DNA strands is because our approach prunes infeasible solutions and does not expand them; therefore a huge part of the solution space is not searched at all. Figure 3 shows the six possible color assignments obtained by our approach.

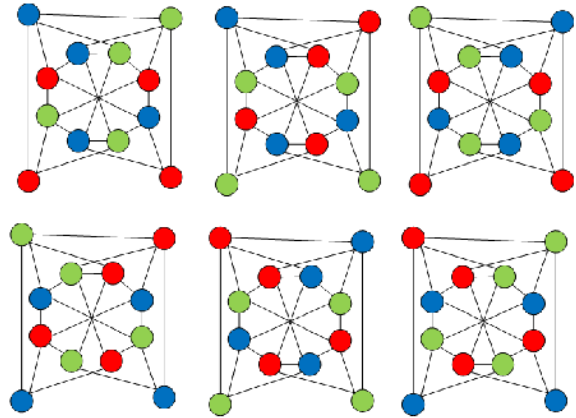

Figure 3. The six possible color assignments obtained by simulating our approach
for the graph shown in Figure 2

## 5. CONCLUDING REMARKS

It has been evidenced that DNA computing is able to solve NP-complete problems in a polynomial time complexity. By using the inherent parallel processing ability of DNA molecules, we can encode all the solution space of the problem and extract feasible solutions by applying polynomial number of biological operations; but the exponential number of DNA strands for encoding all the solution space is still a major problem.

In this paper we proposed an approach based on Adleman Lipton DNA computing model which is able to yield all the solutions to the graph coloring problem without needing the exponential solution space. Our proposed approach has the advantage of discarding a DNA strand before expanding it, when it turns out to be an infeasible solution. This make a remarkable reduction in the number of DNA strands used and makes it possible to apply DNA computing approach for larger graphs than before. We solved a sample graph with 12 vertices by our approach which needed at most 180 number of different DNA strands instead of $3^{12}$ strands. Applying the same technique for solving other NP-complete problems is our future work.